**An injectable, self-healing and MMP-inhibiting hyaluronic acid gel via iron coordination**

Ziyu Gao,[1,2] Xuebin Yang,[1] Elena Jones,[3] Paul A. Bingham,[4] Alex Scrimshire,[4] Paul D. Thornton,[2,*] Giuseppe Tronci[1,5,*]

[1] School of Dentistry, St. James's University Hospital, University of Leeds, UK

[*] E-mail: g.tronci@leeds.ac.uk

[2] School of Chemistry, University of Leeds, Leeds, UK

[*] E-mail: p.d.thornton@leeds.ac.uk

[3] Leeds Institute of Rheumatic and Musculoskeletal Medicine, University of Leeds, UK

[4] Materials and Engineering Research Institute, Sheffield Hallam University, UK

[5] Clothworkers' Centre for Textile Materials Innovation for Healthcare, School of Design, University of Leeds, UK.

**Graphical abstract**

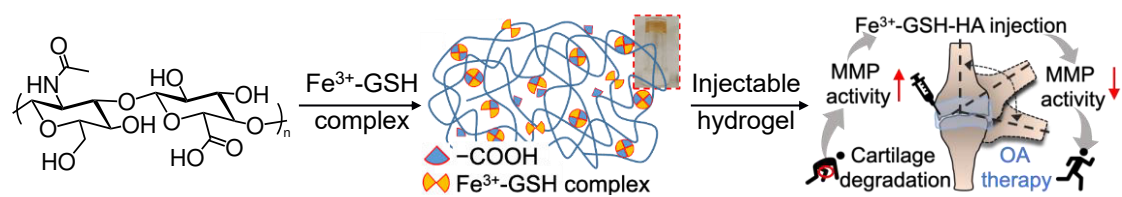

**Abstract**

Regulating the activity of matrix metalloproteinases (MMPs) is a potential strategy for osteoarthritis (OA) therapy, although delivering this effect in a spatially and temporally localised fashion remains a challenge. Here, we report an injectable and self-healing hydrogel enabling factor-free MMP regulation and biomechanical competence *in situ*. The hydrogel is realised within one minute upon room temperature coordination between hyaluronic acid (HA) and a cell-friendly iron-glutathione complex in aqueous environment. The resultant gel displayed up to 300% in shear strain and tolerance towards ATDC 5 chondrocytes, in line with the elasticity and biocompatibility requirements for connective tissue application. Significantly enhanced inhibition of MMP-13 activity was achieved after 12 hours *in vitro*, compared with a commercial HA injection (OSTENIL® PLUS). Noteworthy, 24-hour incubation of a clinical synovial fluid sample collected from a late-stage OA patient with the reported hydrogel was still shown to downregulate synovial fluid MMP activity (100.0±17.6 % → 81.0±7.5 %), with at least comparable extent to the case of the OSTENIL® PLUS-treated SF group (100.0±17.6 % → 92.3±27.3 %). These results therefore open up new possibilities in the use of HA as both mechanically-competent hydrogel as well as a mediator of MMP regulation for OA therapy.

**Introduction**

Osteoarthritis (OA) is a chronic and irreversible disease which results in continuous cartilage degradation, increased joint friction, and pain. The onset and progression of OA is closely linked to proteolytic imbalances, whereby upregulated activity of matrix metalloproteinases (MMPs), particularly MMP-13 (collagenase), results in the pathological breakdown of articular cartilage (Yoshihara et al., 2000) (Burrage et al., 2006) (H. Li et al., 2017). MMP-13 concentration strongly correlates to vascular endothelial growth factor (VEGF) concentration, which plays an important role in angiogenesis and can serve as a biomarker for OA diagnosis and therapeutic monitoring (Kim et al., 2011). In addition, the overexpression of MMP-13 is found in advanced osteoarthritic synovial fluid (Heard et al., 2012). Injectable, non-cytotoxic and biomechanically viable materials that are able to inhibit MMP-13 are highly sought to restore tissue homeostasis and minimise the risks of knee replacement (M. Wang et al., 2013).

Injectable materials enable the delivery and localisation of therapeutic compounds at a target diseased site. In particular, injectable materials that mimic the features of the extracellular matrix (ECM) are ideal therapeutic scaffolds since they enable cell attachment, proliferation and temporally controlled mechanical function with minimal toxic effect following degradation (Stevens & George, 2005) (Blache et al., 2020). As such, they have been widely employed as carriers for improved mesenchymal stem cell (MSC) delivery for bone repair and OA management (M. Liu et al., 2017). Hydrogel systems that contain synthetic polymers have shown promise as materials for OA management due to their injectability and versatility in presenting bioactive functionalities that downregulate MMP activity and prolong the activity of encapsulated MSCs (Clark et al., 2020). Yet, the limited degradability of many synthetic polymers and the demands of polymer synthesis make their translation to commercial products challenging. The design of injectable hydrogels from ECM-derived polymers that can correct proteolytic imbalances may provide an alternative cell-free and regulatory-friendly strategy for OA management, which avoids non-biodegradable synthetic polymers.

Hyaluronic acid (HA) is an anionic non-sulfated glycosaminoglycan that constitutes one of the main components of cartilaginous ECM (Slepecky, 1967). Due to its polysaccharide backbone, a great deal of attention has been put into investigating HA functionalisation for targeted applications, aiming to accomplish tuneable physicochemical properties (Zamboni et al., 2020) and improved cell viability (Zamboni et al., 2017). However, many commercially available HA-based products are in the form of injectable materials, for instance OSTENIL® PLUS, which is routinely applied in the clinic for the treatment of osteoarthritic joints. Significantly improved knee function and pain relief were confirmed through the Visual

Analog Scale (VAS) score and the Western Ontario and McMaster Universities Osteoarthritis Index (WOMAC) score (Kotevoglu et al., 2006)(Dernek et al., 2016). HA injections are usually suggested to be delivered every 1-2 weeks to the joint cavity, although they are unable to control OA-related MMP upregulation. Despite HA's capability to interact with and stimulate chondrocytes *in vivo*, these products are only designed to offer a palliative, short-lived biomechanical solution that is used as a last resort prior to joint replacement. Intelligent HA formulations that include therapeutics for OA treatment through MMP-13 inhibition, and retain mechanical stability, are highly sought. To pursue this vision, a cell-friendly iron-glutathione ($Fe^{3+}$-GSH) complex recently reported by our group (Gao et al., 2020) was investigated for use as both a crosslinker of HA to yield an injectable hydrogel, and as a potential therapeutic to inhibit MMP-13 activity, exploiting the competitive metal-coordinating reaction between thiol complexed iron ($Fe^{3+}$) and active MMPs.

Although some effort afforded the creation of HA-containing gels via metal coordination, e.g. INTERGEL™, unpleasant side-effects and serious complications experienced by many patients call for new safer alternatives (Tang et al., 2006). To prevent tissue damage from ·OH and peroxy-type radicals, which could be generated during hyaluronic acid degradation (Katarina Valachová et al., 2016; Katarína Valachová et al., 2015), it is important to involve reductive components into the HA-based therapeutic material, for example, thiol groups (Katarína Valachová et al., 2015). In this case, introducing cell friendly $Fe^{3+}$-GSH complex into HA hydrogels is worth investigating.

Hydrogel injectability has been pursued via dynamic covalent chemistries in biopolymer-based hydrogels for tissue engineering, including Schiff-base reactions (Huang et al., 2016; S. Li et al., 2020), Diels-Alder reactions (DA) click coupling reactions (Hu et al., 2019) (Spicer, 2020), as well as via thermal gelation mechanisms (Zhang et al., 2019; Lee et al., 2020) compliant with injection-mediated delivery. On the one hand, the formation of covalently crosslinked hydrogels with appropriate mechanical properties in physiological conditions to reduce joint friction has up to now proven challenging. This is largely due to the fact that the presence of covalent crosslinks reduces hydrogel's dynamic tensile, compressive and shear strain, limiting hydrogel's ability to bear multiple load-bearing cycles, as in the case of articular cartilage. On the other hand, although thermosensitive polymer formulations have been developed, only a limited number have been made with HA formulations free of the synthetic polymer phase (Zhang et al., 2019).

Other than covalent networks, redox-based self-healable and injectable polymer hydrogels were achieved that can withstand relatively high shear strain (~50 %) (Chen et al., 2019) (L. Liu et al., 2019). Likewise, metal-coordinated hybrid materials have been reported serving as electroconductive materials (Shi et al., 2015), catalyst supports (Loynachan et al., 2019), and for magnetic resonance imaging (Paquet et al., 2011) (H. Wang et al., 2019). Ultimately, composite hydrogels have been made of multiple biopolymers and bioglass and ionically crosslinked by calcium dications (Yu et al., 2019). The composite material is able to withhold quercetin, an MMP inhibitor, so that 70% reduction in MMP-13 expression was reported after 48 hours, which proved key to induce cartilage repair after 12 weeks *in vivo*. These studies provide novel design concepts that harness the functionalities of metals and peptides, aiming to build simple ECM mimetics with flexible mechanical properties and MMP inhibition capability.

In this work, the straightforward creation of a non-toxic HA-based hydrogel that is injectable and self-healing is reported. HA combined with an iron ($Fe^{3+}$)-glutathione ($Fe^{3+}$-GSH) complex results in the formation of a physical hydrogel upon co-injection. We hypothesised that hydrogel-induced MMP inhibition was accomplished by harnessing the metal-coordinating reaction between thiol-complexed iron ($Fe^{3+}$) and active MMPs. Crucially, the $Fe^{3+}$-GSH complex has the dual function of being the crosslinker within the hydrogel, and also providing a therapeutic effect for inhibiting MMP activity, as confirmed with synovial fluid clinical samples collected from patients with late-stage OA. Consequently, the hydrogel may act as a self-healable scaffold that reduces joint friction and halts cartilage degradation, whilst boosting local cell function. Delivery of this system *in situ* has significant potential in OA therapy, aiming to prevent the degradation of cartilage whilst correcting growth factor concentrations and cellular activity towards cartilage repair.

## 2. Materials and methods

The hyaluronic acid sodium salt (molecular weight: 1,200 kDa, cosmetic grade) was purchased from Hollyberry Cosmetic. L-glutathione (reduced) was purchased Alfa Aesar. Alamar Blue assay kit was from ThermoFisher Scientific. Human recombinant Pro-MMP 13 was purchased from Antibodies.com, and the MMP activity assay kit (Fluorometric Green, ab112146) was from ABChem. All the other reagents were provided by Sigma-Aldrich.

**Rheology of HA solutions supplemented with $Fe^{3+}$-GSH**

Different concentrations of $Fe^{3+}$-GSH complex were added to the HA solution (**Table S1**) to achieve the optimal, most stable, hydrogel. To exclude the influence of HA concentration on gel formation, the final concentration of HA in the gel-forming mixture was controlled to 1.33 wt.% by addition of deionised water. All test group samples were named as "Fe xxx", in which "xxx" corresponds to the volume of $Fe^{3+}$-GSH solution (μL) in the HA solution (mL). All control samples were named as "Ctrl xxx", in which "xxx" corresponds to the volume (μL) of $Fe^{3+}$-GSH solvent (120 mM HCl) per mL of HA solution. The $Fe^{3+}$-GSH-supplemented HA solution was injected onto an MCR 302 Rheometer (Anton Paar) and pressed by a 25 mm parallel plate (1.5 mm gap) at 37 °C with a variable shear rate to study the viscosity of hydrogels formed with different $Fe^{3+}$-GSH complex content.

**Preparation of Fe$^{3+}$-GSH self-healing HA hydrogel (Fe 300)**

The Fe$^{3+}$-GSH complex was prepared using our previous method (Gao et al., 2020). Briefly, 123 mg (0.4 millimoles) of GSH was added to 4 mL FeCl$_3$ aqueous solution (0.1 M), and the mixture was mildly agitated by vortex mixing for 2 min until the solution became yellow. Then, the complex was precipitated by adding 40 mL of ethanol (×3) and collected by centrifugation at 10,000 rpm for 15 min. The Fe$^{3+}$-GSH complex was dried at 37 °C for further use.

10 mg of Fe$^{3+}$-GSH complex was dissolved in 1 mL HCl solution (120 mM). Each 300 μL Fe$^{3+}$-GSH complex solution was added to 1 mL hyaluronic acid solution (2 wt.%) and stirred at room temperature for 1 min to obtain a self-healing hydrogel (Fe$^{3+}$-GSH gel). The self-healing behaviour of all hydrogels formed was characterised by determining the reversible viscosity from a low shear strain (0.01 %) for 200 s, followed by a high shear strain (500 %) measurement for 100 s at 37 °C. The testing frequency was fixed at a constant value of 5 rad·s$^{-1}$. Ten low-to-high shear strain cycles were measured in this process using an Anton Paar MCR 302 rheometer.

**Determination of hydrogel shear modulus and shear strength**

The shear modulus (storage modulus G' and loss modulus G'') of the Fe$^{3+}$-GSH crosslinked hydrogel (Fe 300) was measured via a frequency sweep using an MCR 302 rheometer (Anton Paar). This method was set with a 25 mm parallel plate at 37 °C, 1.5 mm gap, from 1-100 rad/s under 5 % amplitude. G' and G'' were determined at 37 °C over a shear strain range of 0-500 % with a constant angular frequency (5 rad·s$^{-1}$). Every 1.0 mL volume of Fe$^{3+}$-GSH gel was injected onto the sample plate and slightly pressed by a 25 mm parallel plate geometry with a gap of 1.5 mm. Hyaluronic acid with the same amount of HCl solution only was measured as a control for both shear modulus and shear strain.

**Molecular mechanism study**

$^{57}$Fe Mössbauer spectroscopy was applied to study iron chelation and valence. Measurements were carried out using acrylic absorber discs (area: 1.8 cm$^2$) loaded with a dried gel sample to achieve a Mössbauer thickness of 1. The 14.4 keV γ-rays were supplied by the cascade decay of 25 mCi $^{57}$Co in Rh matrix source, oscillated at constant acceleration by a SeeCo W304 drive unit, detected using a SeeCo 45431 Kr proportional counter operating with 1.745 kV bias voltage applied to the cathode. All measurements were carried out at 293 K over a velocity range of ±6 mm·s$^{-1}$, and were calibrated relative to α-Fe foil. Spectral data were fitted using the Recoil software package, using a single Lorentzian line shape necessitated by the low signal/noise ratio obtained for the sample (indicative of its low Fe content).

**Cellular tolerability study**

ATDC 5 chondrocytes were cultured (37 °C, 5% $CO_2$) in a mixed medium of Dulbecco's modified Eagle's medium (DMEM) and Ham's F12 medium (1:1 in volume), supplemented with 5% fetal bovine serum (FBS), and 1 % penicillin-streptomycin. A defined amount of self-healing gel was transferred into individual wells of a 96-well-plate and diluted by cell culture medium to a final concentration of 0 μL (tissue culture plastics, TCPs), 5 μL, 10 μL, 20 μL, 30 μL, 40 μL and 50 μL per well, followed by addition of 100 μL cell suspension ($5\times10^4$ cells/mL) in each (n=4). The cell viability was quantified by Alamar blue assay after 1-day, 3-day, 5-day culture. Cells cultured on TCPs were set as the control group.

**MMP-13 inhibition study with MMP-13–supplemented solution**

The self-healing gel, as well as an HA solution and a commercial HA gel for OA injection, OSTENIL® PLUS (both with the same HA concentration as the self-healing gel), were added to deionised water (×4). Then, 20 μL of each sample was added to individual wells of a 96-well plate, followed by adding 80 μL $H_2O$ per well. Pro-MMP 13 was activated following the manufacturer protocol. Briefly, 5 μL MMP-13 (10 μg MMP-13/20 μL sample) was dissolved in a *p*-aminophenyl mercuric acetate (AMPA) working solution (1 mM) to 1 μg/mL and then incubated at 37 °C for 40 min. Activated MMP-13 was diluted with AMPA solution (2 mM) to 25 ng/mL and then immediately added into the sample wells (each containing 100 μL of the sample), corresponding to a final MMP-13 concentration of 12.5 ng/mL to cover the enzymatic concentration (6 ng/mL) recorded in synovial fluid samples of advanced OA patients (Heard et al., 2012). Deionised water with an equal volume of APMA solution (2 mM) was set as the blank, and deionised water with an equal volume of activated MMP-13 was set as the none treatment group. After 12-hour or 24-hour incubation, MMP-13 activity was quantified via fluorometric assay (Fluorometric Green, ab112146, Abcam) (Liang et al., 2018). 50 μL of each sample was pipetted into a new 96-well-plate, followed by 50 μL of MMP Green Substrate working solution. MMP 13-activity was recorded in fluorescence after 1-hour reaction in dark at 37 °C using a microplate reader (Thermo Scientific Varioskan® Flash, Ex/Em=490/525 nm).

**MMP-13 regulation study with patient collected synovial fluid**

Synovial fluid (SF) samples were collected from late-stage osteoarthritic patients at Chapel Allerton Hospital (Leeds, UK) under ethical approval granted by the National Research Ethics Committee (ethical approval number: 07/Q1205/27). SF samples were stored at -80 ˚C until use. A fluorometric assay kit (Fluorometric Green, ab112146) was used to measure the total proteolytic activity in both SF and hydrogel-incubated SF samples. SF samples were diluted with the MMP assay buffer (×4), and the final $Fe^{3+}$-GSH crosslinked gel dose was increased (×4). 50 μL of diluted SF were mixed with 40 μL of $Fe^{3+}$-GSH crosslinked gel, and 10 µL of deionised water was supplemented in each well to achieve a final concentration of 100 µL/mL [$Fe^{3+}$-GSH crosslinked gel/solution]. The fluorometric assay was conducted after 24-hour incubation following the same assay protocol reported for MMP-13 activity measurement.

**Statistical analysis**

All the samples were tested with at least three replicates (n≥3) and presented as Mean±SD. Statistical significance level was calculated through one-way ANOVA with a p-value at 0.05. Final statistical results were presented as *p≤ 0.05, **p≤ 0.01, ***p≤ 0.001, ****p≤0.0001.

## 3. Results and discussion

Attempts to create hydrogels from HA (2 wt.%) and varying amounts of the $Fe^{3+}$-GSH complex (10 mg/mL) were conducted, and the optimal hydrogel was formed from 300 µL $Fe^{3+}$-GSH complex (10 mg/mL) and 1 mL HA solution (2 wt.%).

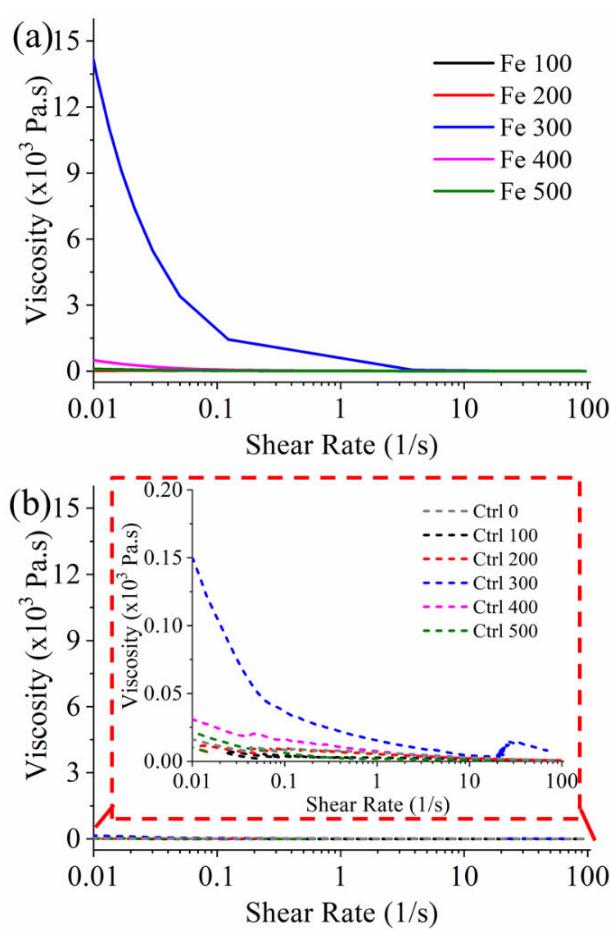

**Fig. 1** Flow curve of aqueous solutions supplemented with (a) either varied $Fe^{3+}$-GSH complex/HA ratio or (b) varied concentration of HA, enlarged within the red box.

A significant decrease in viscosity was observed with increasing shear rate from 0.01 Hz (14,400 Pa·s) to 4 Hz (37 Pa·s), whereas the viscosity remained constant at shear rates between 4 Hz and 100 Hz (**Fig. 1a**). Compared with the other materials created, the stability in hydrogel viscosity suggested a balanced coordination at a $Fe^{3+}$-GSH crosslinker concentration of 300 μL per mL of HA solution. On the other hand, in the HA solution control groups, replacement of the Fe-GSH complex with the HCl solution resulted in significantly lower viscosity(**Fig. 1b**), whereby no significant viscosity variation was observed across the control groups.

The iron oxidation state in the optimal hydrogel (Fe 300) was ferric ($Fe^{3+}$) occupying octahedral coordination (Dyar et al., 2006)) ((Khalil et al., 2013), as determined by $^{57}$Fe Mössbauer spectroscopy (**Fig. 2**), which also confirmed the chelation of $Fe^{3+}$ to HA.

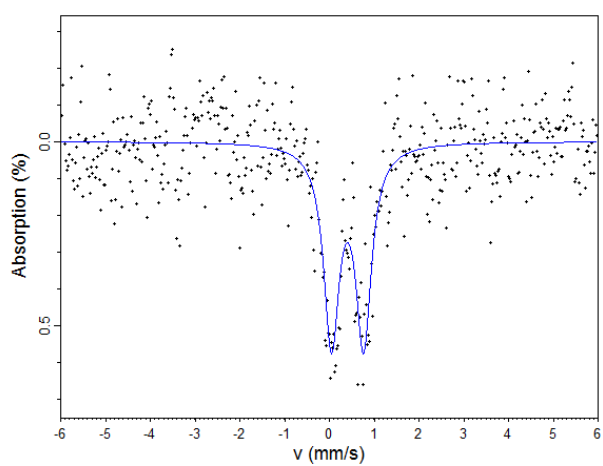

**Fig. 2** Fitted $^{57}$Fe Mössbauer spectrum of dry $Fe^{3+}$-GSH gel at 293 K, relative to thin α-Fe foil. The clear presence of a doublet attributable to paramagnetic $Fe^{3+}$ can be observed, despite the low signal/noise ratio due to the low abundance of $Fe^{3+}$-GSH content in the gel. Fitted centre shift (δ) = 0.41 ± 0.02 mm s$^{-1}$ and quadrupole splitting (Δ) = 0.72 ± 0.02 mm s$^{-1}$ with HWHM linewidth = 0.21 ± 0.02 mm s$^{-1}$.

The confirmed $Fe^{3+}$ state in the hydrogel therefore speaks against a GSH-induced reduction to $Fe^{2+}$ and the consequent generation of toxic reactive oxygen species, supporting the safe injectability of the HA hydrogel in the OA site. In light of these characteristics, the aforementioned hydrogel Fe 300 was chosen for further investigation.

A much higher G' value (120 Pa) was recorded for the Fe 300 gel that contained the $Fe^{3+}$-GSH crosslinker, compared to the HCl-HA control (10 Pa), again indicating that Fe-coordination to HA enables gel formation. Constant storage (G'= 120 Pa) and loss (G''= 70 Pa) moduli of the self-healing gel were successfully measured in frequency sweep mode, confirming a predominantly elastic behaviour in the range of 1-40 rad·s$^{-1}$, whilst the material elasticity was found to decrease at the increased angular frequency (**Fig. 3a**). Although the storage modulus is reduced compared to the chemically crosslinked HA hydrogel (G'=300 Pa), the elastic range was much greater (angular frequency: 1-10 rad·s$^{-1}$) compared to the latter care (Gao et al., 2019). This behaviour illustrates the homogeneous nature of the gel. Conversely, the HCl-HA control sample presented an obvious decrease in moduli from high to low frequency (**Fig. 3b**).

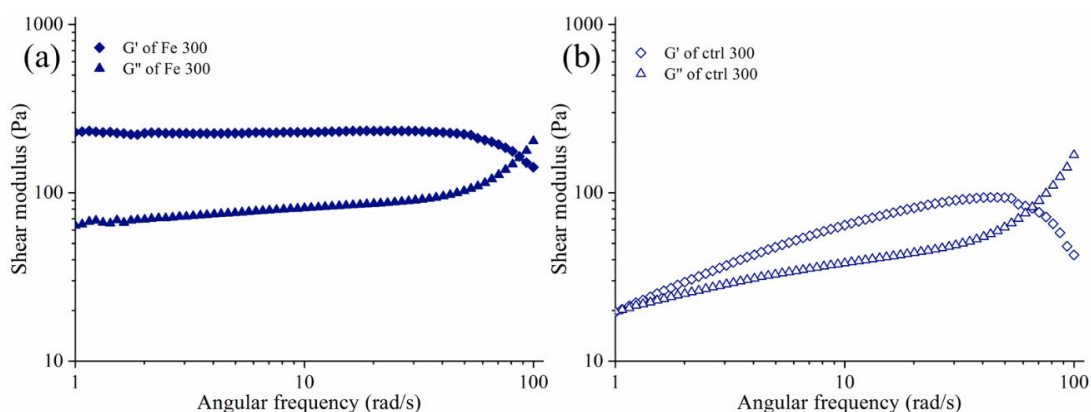

**Fig. 3** Shear modulus of $Fe^{3+}$-GSH hydrogel (a) and ctrl 300 samples (b) recorded during the frequency sweep.

**Fig. 4** reveals the variability of dynamic shear modulus under shear strain (0.01-500 %) for the $Fe^{3+}$-GSH crosslinked gel. A predominantly elastic gel response was observed up to 300 % shear strain, whereby both the storage and loss moduli remained constant when up to 80 % shear strain was applied with 5 rad/s (0.8 Hz) frequency.

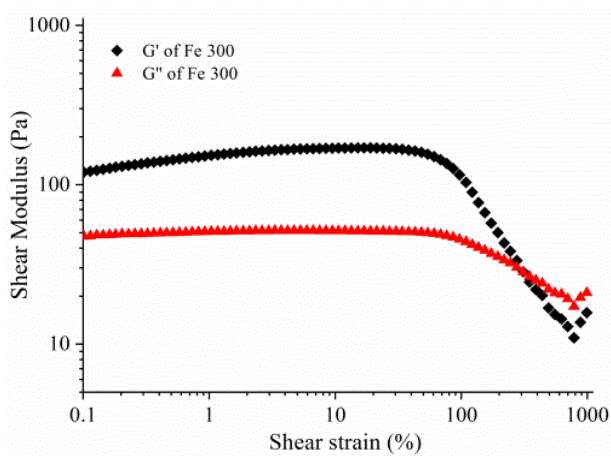

**Fig. 4** Shear modulus of $Fe^{3+}$-GSH gel measured via strain sweep.

These results demonstrate mechanical compliance of the hydrogel with the ranges of shear strain (up to 1 %) and frequency (0.5-2.0 Hz) observed *in vivo* in both connective and fatty tissues (Yoo et al., 2011). In line with previous results, the storage modulus of the $Fe^{3+}$-GSH coordinated gel was found to be greater (105 Pa) than that of the hyaluronic acid control (70 Pa, **Fig. S1**), demonstrating increased mechanical competence.

After 10 cycling tests from low shear strain to high strain, $Fe^{3+}$-GSH crosslinked gels presented a stable complex viscosity in the range of 37-42 Pa·s and 12-16 Pa·s, respectively (**Fig. 5 blue**). This dynamic reversible property confirms that $Fe^{3+}$-GSH crosslinked gels are self-healing materials. The profound degradability of $Fe^{3+}$-GSH crosslinked hydrogel in aqueous solution was confirmed by the decreased viscosity to 0.1-10 Pa·s after being incubated at 37 °C for 5 days (**Fig. 5 grey**).

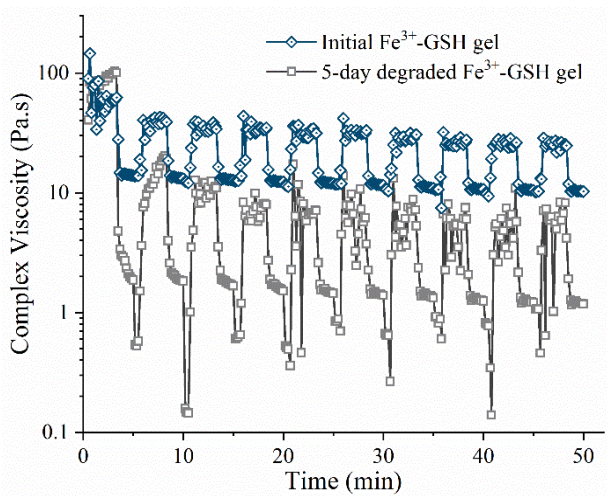

**Fig. 5** Dynamic time-dependent viscosity measurement of the initial (blue) and degraded (grey) $Fe^{3+}$-GSH gel.

The transition from the HA solution to the $Fe^{3+}$-GSH crosslinked self-healing hydrogel was presented in **Fig. 6a&b**. **Fig. 6c** reveals the injectable property of this self-healing hydrogel, and the fact that the material can be absorbed (step 1) by a syringe and then be injected through the syringe tip (step 2), before undergoing extensive elongation (step 3).

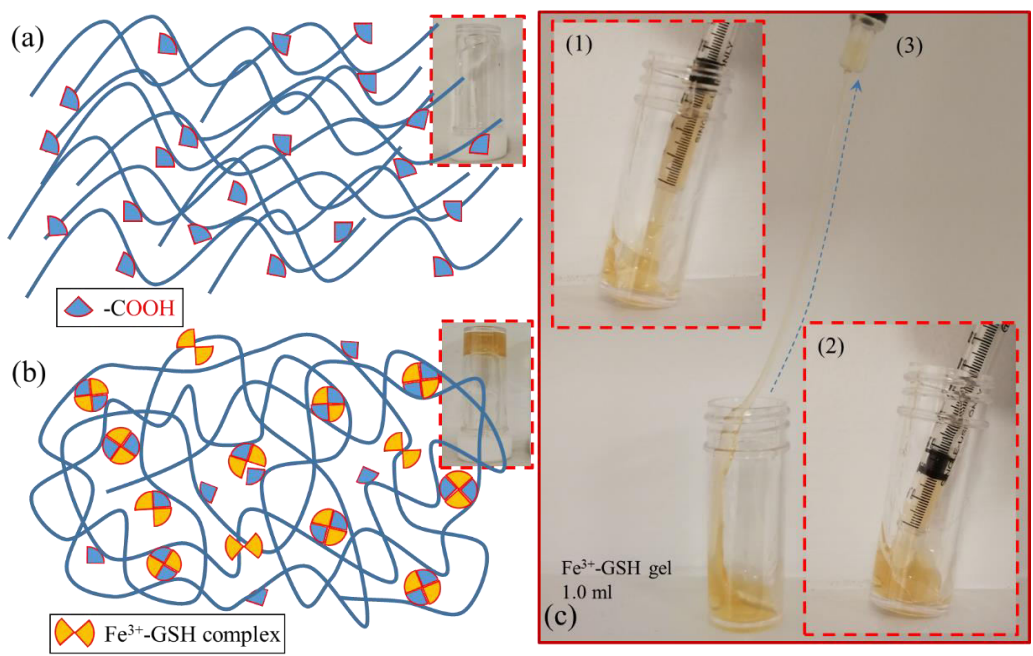

**Fig. 6** Illustration of $Fe^{3+}$-GSH hydrogel formation. (a): Molecular configuration and physical appearance of the HA solution; (b): Proposed coordination structure within, and physical appearance of, the $Fe^{3+}$-GSH hydrogel. (c): Macroscopic properties of $Fe^{3+}$-GSH gel, being loaded up (step 1), injected (step 2) and stretched (step 3).

We could also observe the sticky property of this self-healing hydrogel in step 3; in line with previous viscosity analysis, the adhesive properties of HA were enhanced by $Fe^{3+}$-GSH induction. This feature is key to enable confined application and adhesion of the gel to cartilage, aiming to stabilise the joint cavity and to reduce bone-to-bone friction, which is essential to preserve the cartilage interface (Abubacker et al., 2018).

The dose of $Fe^{3+}$-GSH crosslinked HA gel that is tolerated by ATDC 5 chondrocytes was then determined *in vitro* via Alamar Blue assay (**Fig. 7**).

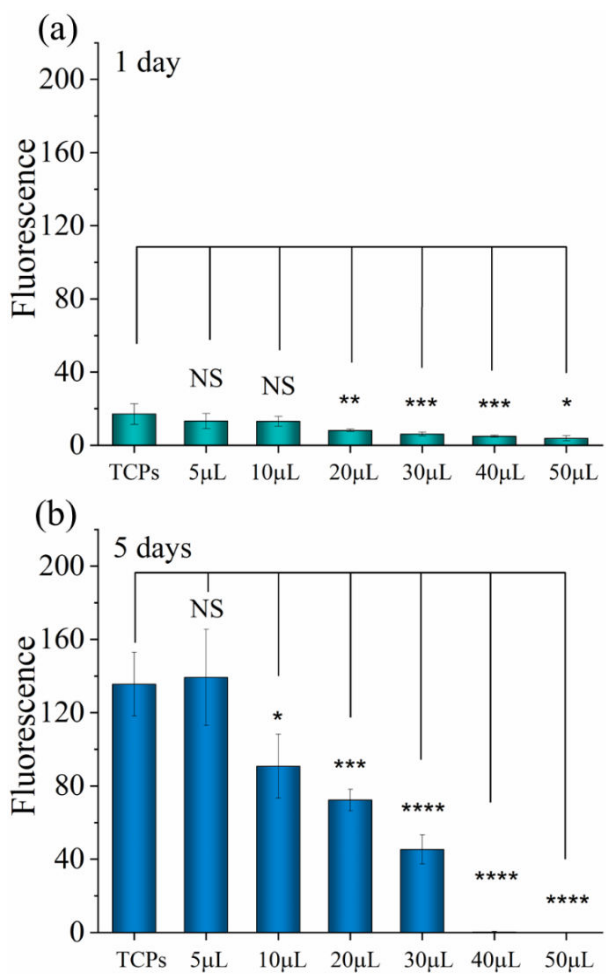

**Fig.7** ATDC 5 cells viability when growing with $Fe^{3+}$-GSH gel after day 1 and 5. No significant differences are labelled with "NS". Significant differences are observed in each group with respect to the TCPs group at the same time point (n=4). * $p < 0.05$, ** $p < 0.01$, *** $p < 0.001$, **** $p < 0.0001$.

As expected, the hydrogel reveals a dose-dependent impact on cellular metabolic activity. At day 1, the lower dose (e.g. 5 and 10 μL) of $Fe^{3+}$-GSH crosslinked HA gel did not show significant effect compared to the case of the TCPs control group ($p > 0.05$). However, the high dose groups (e.g. > 20 μL) significantly reduced the metabolic activity of ATCD-5 cells compared to the control group ($p \leq 0.01$, 0.001, 0.001, 0.05, respectively). Clearly, no significant difference in cellular activity was observed following 1-day cell culture in either TCP or lower doses of $Fe^{3+}$-GSH crosslinked hydrogel (with both 5 μL and 10 μL dose). At day 5, only the 5 µL group was well tolerated (p > 0.05), but all the other higher dose groups (e.g. > 10 µL) were significantly detrimental to the metabolic activity of the cells compared to the control group (p ≤ 0.05, 0.001, 0.0001 respectively). Furthermore, the increase in metabolic activity recorded from day 1 to day 5 in ATDC 5 cells cultured with 5-30 µL hydrogel (**Table 1**) was similar to that measured in cells treated with the TCPs control group (7.9 times). This observation indicates that decreased doses (e.g. ≤ 30 µL) of $Fe^{3+}$-GSH hydrogel did not affect the cell proliferation (e.g. cell doubling) in this time window, in contrast to the case where higher doses (e.g. ≥ 40 µL) were applied. Given that the initial cell seeding density (5,000 cells per well) was maintained across all hydrogel groups (5-50 µl), the reduced cellular metabolic activity observed with increased gel volume (> 30 µl) is likely attributed to the relatively small number of cells cultured with increased sample dosages.

**Table 1** Variation in ATDC 5 cellular activity over 5-day culture with varied hydrogel dosage.

| Hydrogel dosage | Average cellular activity increase |
|---|---|
| 0 µl (TCP)[*] | 7.9 |
| 5 µL | 10.5 |
| 10 µL | 7.0 |
| 20 µL | 8.8 |
| 30 µL | 7.4 |
| 40 µL | 0.1 |
| 50 µL | 0 |

[*] Cells cultured in hydrogel-free Tissue Culture Plastic (TCP).

This observation may suggest that the gels under 30 µL dose were temporarily toxic after 1-day. However, the proliferation of the remaining ATDC 5 cells was not affected, an explanation which is supported by the optical microscope images of cells cultured for 1 (**Fig. S2**) and 5 days (**Fig. S3**). In contrast, no cellular tolerability was observed in both 40 and 50 µL hydrogel groups over 5 days.

The capability of the $Fe^{3+}$-GSH crosslinked hydrogels to inhibit proteolytic activity was then assessed, whereby MMP-13 was selected as a well-known upregulated protease in late-stage OA. By selecting MMP-13-supplemented aqueous solutions as a defined *in vitro* environment, incubation of $Fe^{3+}$-GSH hydrogel resulted in a reduction of MMP-13 activity after 12 hours (95.7±3.4 %). A significant reduction in MMP-13 activity (92.9±1.4 %) was recorded after 24 hours, compared to the positive control group (p<0.001) **(Fig. 8)**. On the other hand, no significant activity difference was observed between MMP-13-supplemented solutions and the same solutions following incubation with either soluble, complex-free GSH (103.1±7.6 %) (Gao et al., 2020) or native HA after 24 hours (98.5±5.0 %). In OSTENIL® PLUS, no reduction in MMP-13 activity was seen after 12 hours, but a significant reduction (p<0.05) in activity was observed after 24 hours (96.1±1.7 %), with respect to the pristine MMP-13 solution. A comparison between the $Fe^{3+}$-GSH crosslinked gel and OSTENIL® PLUS reveals that increased MMP-13 inhibition occurred in the presence of the $Fe^{3+}$-GSH crosslinked hydrogel after 12 hours (p<0.01), which was maintained after 24 hours (p<0.05).

These results provide indirect evidence that the hydrogel-induced MMP-13 inhibition was achieved via chelation of respective iron sites with active MMPs, rather than by complexation of the free zinc sites of active MMPs (Liang et al., 2018) with HA's carboxylic groups on the other hand. These observations support the key role played by the $Fe^{3+}$-GSH complex in both hydrogel crosslinking and MMP inhibition.

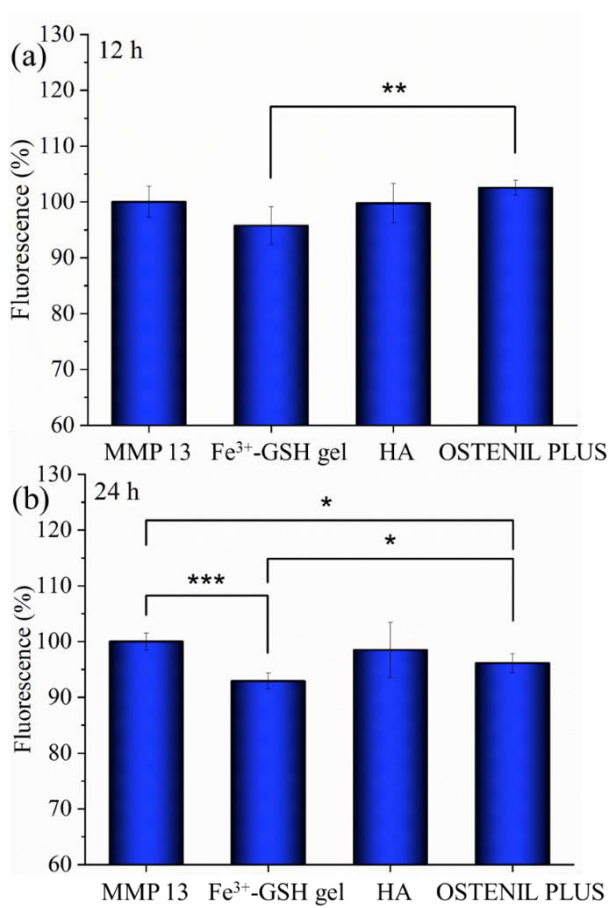

**Fig.8** Variation of MMP-13 activity in MMP-13–supplemented solutions after 12-hour (a) and 24-hour (b) incubation with either the $Fe^{3+}$-GSH crosslinked hydrogel, an HA solution or the OSTENIL® PLUS commercial injection. Data are presented as Mean ± SD, statistical analysis was carried out between each two groups and labelled as *p≤ 0.05, **p≤ 0.01, ***p≤ 0.001, otherwise means no significant difference at p=0.05 level.

A sample of synovial fluid (S162) collected from patients with late-stage OA was used to investigate the MMP-regulating capability of the $Fe^{3+}$-GSH crosslinked gel in near-physiologic conditions, and to further corroborate the previous findings obtained for hydrogel-mediated MMP-13 inhibition in a defined *in vitro* environment, as the overall proteolytic activity, including MMP-1, -2, -3, -7, -8, -9 and -13, were confirmed to have increased activity in advanced OA (Yoshihara et al., 2000). **Fig. 9** reveals that lower overall MMP activity and smaller standard deviations were observed for the $Fe^{3+}$-GSH crosslinked gel (81.0±7.5 %) compared to the native SF group (100.0±17.6 %), with a p-value of 0.0942. Although OSTENIL® PLUS presented a lower average value of activity (92.3±27.3 %) compared to native SF (p=0.6528), a larger standard deviation was recorded for this group versus both SF and the $Fe^{3+}$-GSH crosslinked gel.

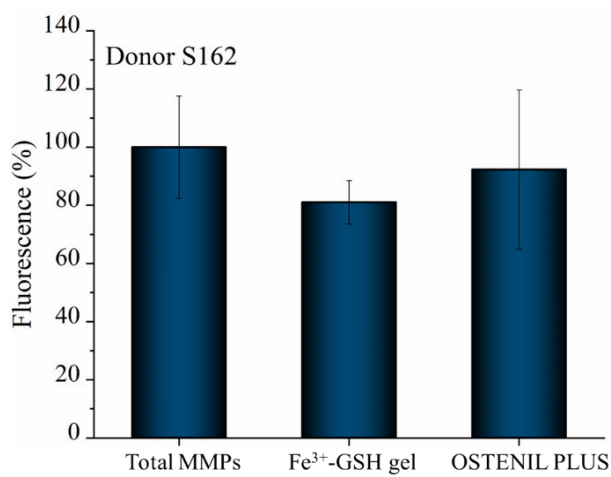

**Fig.9** Variation of MMP activity recorded in a patient collected SF sample after 24-hour incubation with either the $Fe^{3+}$-GSH crosslinked hydrogel or the OSTENIL® PLUS commercial injection (n=4). The SF sample was collected from a patient (donor S162) with late-stage OA.

The results obtained with the clinical SF sample in the absence of MMP activating reagents, i.e. APMA, therefore confirm the new MMP inhibition functionality introduced in the $Fe^{3+}$-GSH crosslinked hydrogel. These results therefore support the use of this material as both a mechanically-competent hydrogel and as a mediator of MMP regulation for OA therapy. The confirmation of hydrogel performance with patient collected samples also lay down new possibilities on the use of human synovial fluid for the preclinical evaluation of medical devices intended for osteoarthritis management, yet minimising reliance on animal testing.

**Conclusions**

A drug-free $Fe^{3+}$-GSH crosslinked injectable hydrogel was prepared with integrated self-healing and MMP inhibition functionalities. The coordination mechanism to yield the hydrogel was confirmed by shear frequency sweep tests, which revealed a storage modulus more than ten times higher than the loss modulus. $^{57}$Fe Mössbauer spectroscopy revealed that Fe was present in the hydrogel as octahedrally-coordinated $Fe^{3+}$, so that risks of $Fe^{2+}$-mediated ROS generation and ROS-mediated toxicity were minimised, supporting the hydrogel applicability in biological environment. The hydrogel could hold up to 300% shear strain and presented a stable complex viscosity (37-42 → 12-16 Pa·s) after 10 cycling tests from low to high strain. *In vitro*, the gel proved to be well tolerated by ATDC 5 chondrocytes and to support cell proliferation during a five day-culture. Furthermore, the gel demonstrated the inhibition of MMP activity after 24 hour-incubation in both an MMP-13–supplemented aqueous solution and a patient collected sample of synovial fluid, in light of the metal-coordinating reaction between thiol-complexed iron ($Fe^{3+}$) and active MMPs. These results therefore demonstrate that the hydrogel's biomechanical competence was successfully integrated with drug-free MMP regulation capability. The simple material design, together with the hydrogel's injectability, and biochemical and self-healing functionalities support further development of this system for drug-free OA therapies.

**Acknowledgements**

PAB acknowledges with thanks funding support from the UK Engineering and Physical Science Research Council (EPSRC) under Grant EP/R020957/1, New Industrial Systems: Manufacturing Immortality.

**Supporting information**

**Table S1** Formulation of different samples

| Sample name | HA (2 wt.%, mL) | $Fe^{3+}$-GSH (10 mg/ml, μL) | HCl (120 mM, μL) | $H_2O$ (μL) |
|---|---|---|---|---|
| Fe 100 | 1 | 100 | 0 | 400 |
| Fe 200 | 1 | 200 | 0 | 300 |
| Fe 300 | 1 | 300 | 0 | 200 |
| Fe 400 | 1 | 400 | 0 | 100 |
| Fe 500 | 1 | 500 | 0 | 0 |
| Ctrl 0 | 1 | 0 | 0 | 500 |
| Ctrl 100 | 1 | 0 | 100 | 400 |
| Ctrl 200 | 1 | 0 | 200 | 300 |
| Ctrl 300 | 1 | 0 | 300 | 200 |
| Ctrl 400 | 1 | 0 | 400 | 100 |
| Ctrl 500 | 1 | 0 | 500 | 0 |

*Solvent information: HA ($H_2O$), $Fe^{3+}$-GSH (120 mM HCl), HCl ($H_2O$).*

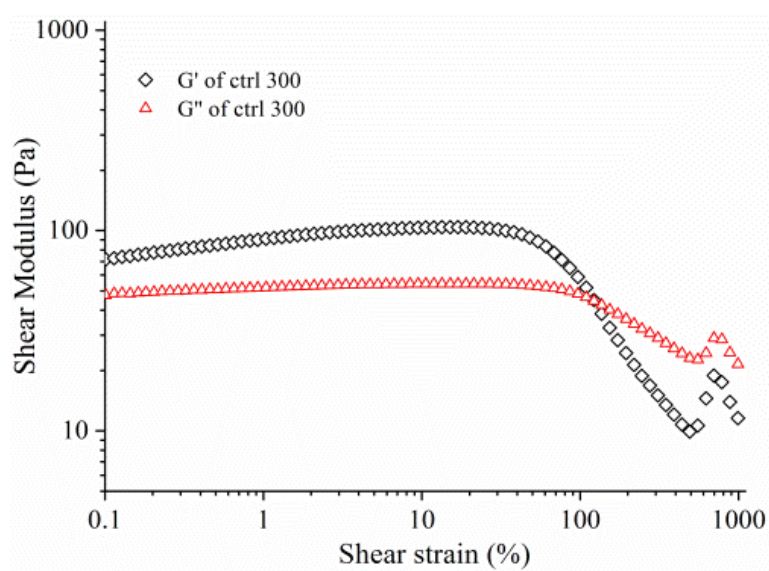

**Fig.S1** Shear modulus of HA-HCl control sample measured via strain sweep.

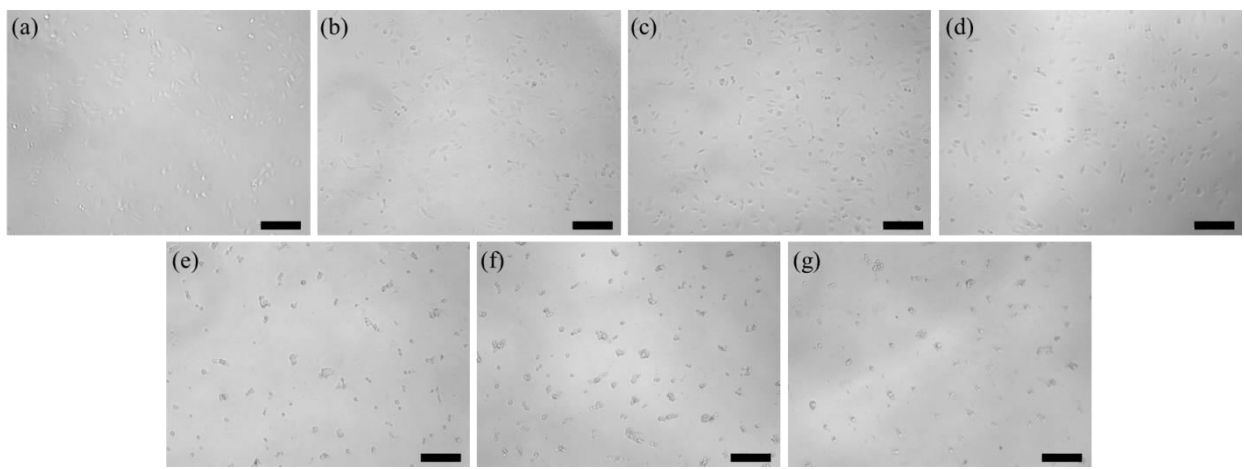

**Fig.S2** Optical images of ATDC 5 cells after 1-day culture with $Fe^{3+}$-GSH gel, (a) 0 μL (TCPs), (b) 5 μL, (c)10 μL, (d) 20 μL, (e) 30 μL, (f) 40 μL and (g) 50 μL per well. Scale bar: 100 μm.

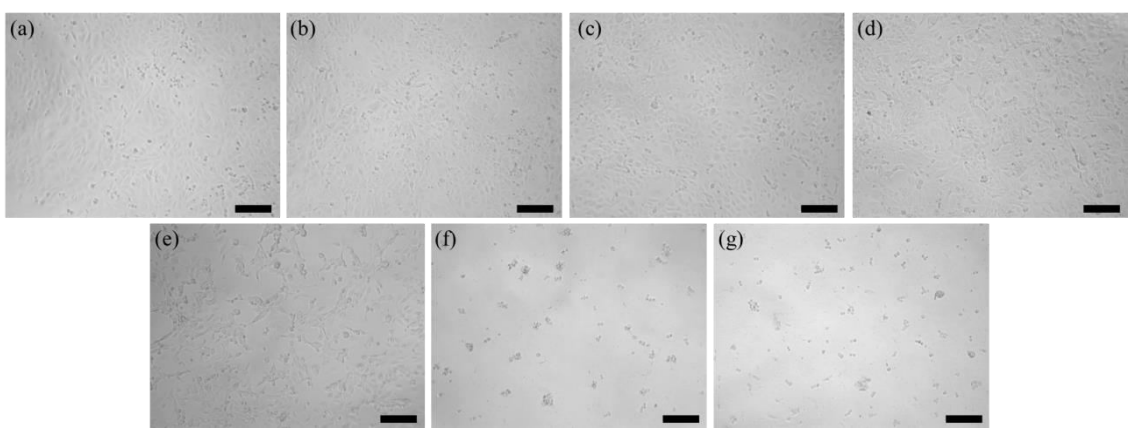

**Fig.S3** Optical images of ATDC 5 cells after 5-day culture with $Fe^{3+}$-GSH gel, (a) 0 μL (TCPs), (b) 5 μL, (c)10 μL, (d) 20 μL, (e) 30 μL, (f) 40 μL and (g) 50 μL per well. Scale bar: 100 μm.